\documentclass{article}
\usepackage{spconf,amsmath,graphicx}
\usepackage{adjustbox}
\usepackage{multirow}
\usepackage{multicol}
\usepackage{amssymb}
\usepackage{url}
\newlength{\bibitemsep}
\newlength{\bibparskip}
\let\oldthebibliography\thebibliography
\renewcommand\thebibliography[1]{%
  \oldthebibliography{#1}%
  \setlength{\parskip}{\bibitemsep}%
  \setlength{\itemsep}{\bibparskip}%
  \footnotesize
}

\usepackage{array}
\newcolumntype{L}[1]{>{\raggedright\let\newline\\\arraybackslash\hspace{0pt}}m{#1}}

\title{MULTIMODAL DEPRESSION CLASSIFICATION USING ARTICULATORY COORDINATION FEATURES AND HIERARCHICAL ATTENTION BASED TEXT EMBEDDINGS}
\name{Nadee Seneviratne and Carol Espy-Wilson\thanks{This work was supported by the UMCP \& UMB Artificial Intelligence + Medicine for High Impact Challenge Award and the National Science Foundation grant numbered 2124270. We thank Dr. James Mundt for the depression databases MD-1\&2 \cite{MUNDT2007, MUNDT2012} and Dr. Thomas Quatieri and Dr. James Williamson for granting access to the MD-2 database which was funded by Pfizer.}}
\address{University of Maryland - College Park, USA}
\begin{document}
%
\maketitle
\begin{abstract}
Multimodal depression classification has gained immense popularity over the recent years. We develop a multimodal depression classification system using articulatory coordination features extracted from vocal tract variables and text transcriptions obtained from an automatic speech recognition tool that yields improvements of area under the receiver operating characteristics curve compared to unimodal classifiers (7.5\% and 13.7\% for audio and text respectively). We show that in the case of limited training data, a segment-level classifier can first be trained to then obtain a session-wise prediction without hindering the performance, using a multi-stage convolutional recurrent neural network. A text model is trained using a Hierarchical Attention Network (HAN). The multimodal system is developed by combining embeddings from the session-level audio model and the HAN text model. 
\end{abstract}
%
\section{Introduction}
\label{sec:intro}
Major Depressive Disorder (MDD) is a mental health disorder with serious consequences. Previous studies have shown that vocal biomarkers developed using prosodic, source, and spectral features \cite{CUMMINS201510} can be very effective in automatic depression detection to enable timely diagnosis and prompt treatments. 

MDD is known to cause changes in articulatory coordination of speech due to a neurological condition called psychomotor slowing which is a necessary feature of MDD \cite{Whitwell1937, ManualMentalDisOrd, WIDLOCHER198327}. Articulatory Coordination Features (ACFs) have yielded successful results in distinguishing depressed speech from non-depressed speech by quantifying these changes in the timing of speech gestures \cite{WILLIAMSON2019, Espy-Wilson2019, Seneviratne2020}. Previously, the correlation structure of the formants or Mel Frequency Cepstral Coefficients (MFCCs) was used as a proxy for articulatory coordination to derive indirect ACFs which showed promise in the depression detection task \cite{WILLIAMSON2019}. Authors of this paper showed in their previous work, that by using Vocal Tract Variables (TVs) as a direct measure of articulation to quantify changes in depressed and non-depressed speech can yield relatively better results in the depression detection task \cite{Espy-Wilson2019, Seneviratne2020} and for depression severity level classification task \cite{seneviratne21_interspeech}. TV-based ACFs also showed promise as a robust set of features for depression classification by generalizing well across the two databases \cite{seneviratne21b_interspeech}. 

Deep learning based depression detection is a highly researched area with promising results \cite{Zhao2020, Ma2016}. Among these works, multimodal depression classification and severity prediction attempt to further improve the performance through inter-learning among different modalities \cite{Niu2021,Ray2019, Lam2019, Fan}.  
Several studies have used the approach of aggregating segment-level predictions to obtain a final subject/session-level prediction using techniques such as plurality voting (PV) \cite{Vazquez2020, Aloshban2020}, max-pooling \cite{Harati2021} or recurrent neural networks (RNN) \cite{Yin2019,seneviratne21_interspeech}. 
This is especially useful in a setting where there's a lack of training samples to train a deep learning model using full audio recordings which can lead to overfitting issues. It was shown empirically that the strengths of the segment-level classifier are amplified as a result of its repeated usage in the session-level classifier \cite{Aloshban2020, seneviratne21_interspeech}. In a multimodal setting, most of these aggregating approaches have used one-to-one correspondence among segments from different modalities \cite{Yin2019, Aloshban2020}.

The key contributions of this paper are as follows: 

(1) The development of a multimodal system using depression corpora that contain only speech data. Utilizing ASR to obtain text transcriptions, we show that for the first time, the performance of binary depression classification can be improved by using TV-based ACFs and textual features. 
The generalizability is improved by sourcing data from two different depression databases. The fusion strategy of the proposed architecture enables us to segment data from different modalities independently, in the most optimal way for each modality, when performing session-level classification from segment-level classification.

(2) The analysis of the constraints to be satisfied by the segment-level classifier to yield a stronger session-level classifier. Using a multi-stage convolutional recurrent neural network, these analytical findings are validated empirically using different sets of ACFs and openSMILE features.


\vspace*{-3mm}
\section{Feature Extraction}
\label{sec:audio-feats}
\vspace*{-2mm}
\subsection{Audio Features}
\vspace*{-2mm}
\emph{\textbf{Vocal Tract Variables (TVs):} } Developed based on Articulatory Phonology \cite{Browman1992}, TVs define the kinematic state of 5 distinct constrictors (lips, tongue tip, tongue body, velum, and
glottis) located along the vocal tract in terms of their constriction degree and location. We use a speaker-independent deep neural network based speech inversion system \cite{Sivaraman2019} to estimate 6 TVs for 3 of the constricting organs - Lip Aperture, Lip Protrusion, Tongue Tip Constriction Location, Tongue Tip Constriction Degree, Tongue Body Constriction Location and Tongue Body Constriction Degree. In addition, we use the periodicity and aperiodicity measures obtained from an Aperiodicity, Periodicity and Pitch (APP) detector \cite{Deshmukh2005} to represent the glottal TV. At this time, we provide no information on the velum.

For comparison purposes, we also trained models using ACFs derived from formants and MFCCs.
The first three formant frequencies were obtained using the Karma formant tracking tool \cite{Mehta_2012}.
12 MFCC time series were extracted (window size of 20ms, overlap of 10ms) discarding the (1\textsuperscript{st} MFCC coefficient. We created two more sets of ACFs by appending the same glottal parameters to formants (FMT+GL) and MFCCs (MFCC+GL) to investigate the effect of adding voice source information.

\vspace*{-3mm}
\subsubsection{Articulatory Coordination Features (ACFs)}
\vspace*{-1mm}
ACFs can be used to characterize the level of articulatory coordination and timing. To measure the coordination, assessments of the multi-scale structure of correlations among the time series signals such as TVs were used. 

We use the channel-delay correlation matrix proposed in \cite{Huang2020} as the ACFs in this work. For an $M$-channel feature vector $\mathbf{X}$ (such as TVs or formants), the delayed correlations ($r_{i,j}^d$) between $i^{th}$ channel $\mathbf{x_i}$ and $j^{th}$ channel $\mathbf{x_{j}}$ delayed by $d$ frames, are computed as:
\vspace{-5pt}
\footnotesize
\begin{equation}
    r_{i,j}^d = \frac{\sum_{t=0}^{N-d-1}x_i[t]x_j[t+d]}{N-|d|}
\end{equation}
\normalsize
where N is the length of the channels.
The correlation vector for each pair of channels with delays $d \in [0,D]$ frames will be constructed as follows:
\vspace{-5pt}
\small
\begin{equation}
    R_{i,j} = \begin{bmatrix}r_{i,j}^0, & r_{i,j}^1, & \dots & r_{i,j}^D \end{bmatrix}^T \in \mathbb{R}^{1\times (D+1)}
\end{equation}
\normalsize
The delayed auto-correlations and cross-correlations are stacked to construct the channel-delay correlation matrix:
\small
\begin{equation}
    \mathit{\widetilde{R}_{ACF}} =  \begin{bmatrix}R_{1,1} & \dots & R_{i,j} & \dots & R_{M,M}\end{bmatrix}^T \in \mathbb{R}^{M^2\times (D+1)}
\end{equation}
\normalsize

Information pertaining to multiple delay scales are incorporated into the model by using dilated Convolutional Neural Network (CNN) layers with corresponding dilation factors while maintaining a low input dimensionality. Each $R_{i,j}$ will be processed as a separate input channel in the CNN model.
Before computing the ACFs, feature vectors 
were standardized individually.

\vspace*{-3mm}
\subsubsection{Baseline Acoustic Features}
\vspace*{-2mm}
We trained a baseline model using the openSMILE features (window size of 20ms, overlap of 10ms) to benchmark the performance of models trained using ACFs. The extended Geneva Minimalistic Acoustic Parameter Set (eGeMAPS) \cite{eGeMAPS} was extracted using the openSMILE toolkit \cite{opensmile2010}. This 23-dimensional feature set consists of spectral, cepstral, prosodic and voice quality parameters. 

\vspace*{-3mm}
\subsection{Textual Features}
\vspace*{-2mm}
Linguistic features reveal important information about the mental health of a depressed subject. Therefore adding semantic contextual information should help to improve our models. We used Google speech-to-text API to obtain transcribed text of the Free Speech (FS) recordings that were used to train the audio models. Since the Hierarchical Attention Network (HAN) can be expected to explicitly capture contextual information we decided to use context-independent GloVe word embeddings (100-dimensional) \cite{pennington-etal-2014-glove} to initialize the embedding layer of the text model. 


\section{Segment to Session-Level Classification}
\label{sec:proof}
\vspace*{-2mm}

Let $\mathbf{h}$ denote the segment-level classifier, $\mathbf{H}$ denote the session-level classifier, $\mathbf{C}$ be an oracle which provides ground-truth, \{$C_0,C_1$\} be the classes and $S$ be a session consisting of {$S_1, S_2, ..., S_N$} segments. Let's consider an arbitrary class $C_0$ and prove that the recall of class $C_0$ of a PV session-level classifier is better than that of the segment-level classifier it is based on. We use a PV classifier here for simplicity, however RNN based approaches yield more generalizable classifiers.

We assume \footnotesize $ p_j = \Pr(h(S_i)=C_j \mid C(S_i)=C_0)$ \normalsize to be the same  $\forall i$ and each segment classification is independent. Note that $p_0$ is the recall of class $C_0$ and $p_0 + p_1 = 1$. Consider a PV classifier which breaks ties by randomly selecting one class. Let $P_0$ denote the recall of the combined classifier.

\vspace{-15pt}
\tiny
\begin{multline}
\label{eq:pv}
    P_0 = \Pr(H(S)=C_{0} \mid C(S)=C_0) = \\
    \sum_{k=\lceil N/2 \rceil}^{N} \binom{N}{k} d(k) p_0^k  p_1^{N-k} \text{ where } 
    d(k) = 
    \begin{cases}
        1/2, \text{if } k = N/2\\
        1, \text{\footnotesize otherwise} 
    \end{cases}
\end{multline}
\small

Using $\binom{N}{k} = \binom{N}{N - k}$ and $d(k) = d(N-k)$, the sum of coefficients of \eqref{eq:pv} can be written as,
\tiny
\[
 \sum_{k=\lceil N/2 \rceil}^{N} \binom{N}{k} d(k) =  \sum_{k=\lceil N/2 \rceil}^{N} \binom{N}{N - k} d(N - k) = \sum_{k=0}^{\lfloor N/2 \rfloor} \binom{N}{k} d(k)
\] 
\small
With that we have,
\tiny
\[
  2\sum_{k=\lceil N/2 \rceil}^{N} \binom{N}{k} d(k) =  \sum_{k=\lceil N/2 \rceil}^{N} \binom{N}{k} d(k) +  \sum_{k=0}^{\lfloor N/2 \rfloor} \binom{N}{k} d(k) = \sum_{k=0}^{N} \binom{N}{k}
\]
\small
Therefore we have the sum of coefficients of \eqref{eq:pv},
\tiny
 \[ \sum_{k=\lceil N/2 \rceil}^{N} \binom{N}{k} d(k) = \frac{\sum_{k=0}^{N} \binom{N}{k}}{2} = (1+1)^N /2 = 2^{N-1} \]
 \small

Consider the difference of recalls. We use \eqref{eq:pv} to substitute for $P_0$ and artificially multiply $p_0$ by a term equal to $1$ to ensure that the coefficient sums of both the terms are equal. Since $p_0 + p_1 = 1$,

\vspace{-10pt}
\tiny
\begin{multline}
\label{eq:ineqal}
 P_{0} - p_{0} = \underbrace{\sum_{k=\lceil N/2 \rceil}^{N} \binom{N}{k} d(k) p_0^k  p_1^{N-k}}_\text{\footnotesize Coefficient sum is $2^{N-1}$}
  -  p_{0} {\underbrace{(p_0 + p_1)^{N-1}}_\text{\footnotesize Coefficient sum is $2^{N-1}$}}\\
  = \sum_{k=\lceil N/2 \rceil}^{N} \left[d(k)\binom{N}{k} - \binom{N-1}{k-1}\right] p_0^k  p_1^{N-k}  \\
   - \sum_{k=1}^{\lceil N/2 \rceil - 1} \binom{N-1}{k-1} p_0^k  p_1^{N-k} 
\end{multline}
\small
Given that
\tiny
\[
d(k)\binom{N}{k} - \binom{N-1}{k-1} = \binom{N-1}{k-1} (d(k)N/k - 1) = \begin{cases}
        0, \text{if } k \in \{N/2, N\}\\
        > 0, \text{ \footnotesize otherwise} 
    \end{cases}
\]
\small
we can group the terms in the expansion of \eqref{eq:ineqal} into pairs of the form
\footnotesize
\begin{equation}
  p_0^{r}p_1^{l}(p_0^{t}-p_1^{t}) \text{ where }  r > 0, l \geq 0, t \geq 0 \text{ and } r+l+t = N  
\end{equation}
\normalsize
which are non-negative when $p_0 \geq p_1$. Therefore we establish that  $P_{0} - p_{0} \geq 0$ or $P_{0} \geq p_{0}$ when $p_0 \geq p_1$.
That is, if all the classes have better than 50\% recall in the segment-level classifier, PV based session-level classifier would result in a better recall for all the classes.

\begin{figure}[htb]

\begin{minipage}[b]{1.0\linewidth}
  \centering
  \centerline{\includegraphics[width=\columnwidth]{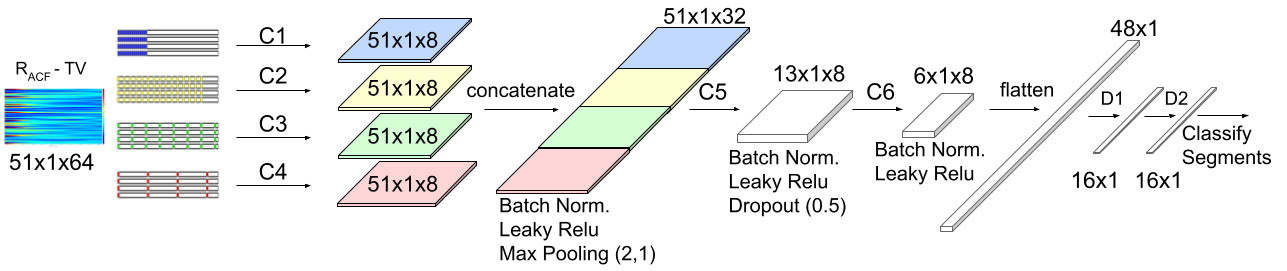}}
  \centerline{(a)}\medskip
\end{minipage}
\begin{minipage}[b]{1.0\linewidth}
  \centering
  \centerline{\includegraphics[width=8cm, height=4.5cm, keepaspectratio]{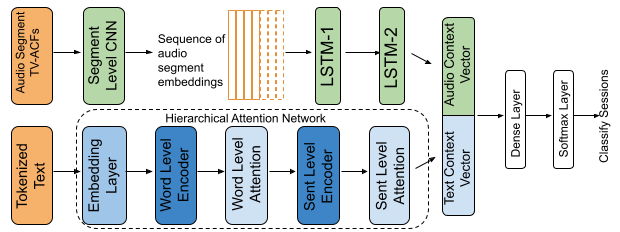}}
  \centerline{(b)}\medskip
\end{minipage}
\vspace*{-8mm}
\caption{(a) Dilated CNN architecture for segment-level classification. Hyper-parameters for the best performing audio model (TV-based ACF) are mentioned. Kernel size of C5 and C6 is (3,1). (b) Architecture of the multimodal classifier. LSTM-1 and LSTM-2 have 128 and 64 hidden units (HU) and 0.7 and 0.7 dropout probabilities (DP), respectively. The word-level encoder and sentence-level encoder have 100 HU each and 0.3 and 0.1 DP, respectively. The Dimension of attention layers are 64. The final Dense Layer (before the Softmax Layer) has 64 HU.}
\label{fig:dnn_model}
\end{figure}

\vspace*{-2mm}
\section{Model Architectures}
\label{sec:models}
\vspace*{-2mm}
\subsection{Audio Model}
\label{sec:audio-model}
\vspace*{-2mm}
\emph{\textbf{Baseline Segment-Level Classifier: }}
We trained a CNN using the openSMILE features as a baseline model for this task. The input is passed through two sequential 1-D (across time axis) convolutional layers. 
Each convolutional layer is followed by batch normalization, leaky ReLU activation, dropouts 
and a max-pooling layer. 
The output from the second max-pooling layer is flattened and passed through two dense layers 
to perform classification at the output layer. The output of the second dense layer is extracted and used as the input to the session-level classifier.

\emph{\textbf{Dilated CNN based Segment-Level Classifier for ACFs: }}
A dilated CNN proposed in \cite{Huang2020} was trained using the ACFs to classify the segments (Fig. \ref{fig:dnn_model}a). The input $\mathit{\widetilde{R}_{ACF}}$ is fed into 4 parallel convolutional layers with different dilation rates $n=\{1,3,7,15\}$ and a kernel size of $(15,1)$ which resembles the multiple delay scales. The outputs of these 4 parallel layers are concatenated and then passed through two sequential convolutional layers. This output is flattened and passed through two dense layers 
to perform segment-level classification in the output layer. All convolutional layers used LeakyReLU activation, whereas the dense layers used ReLU activation with $l_2$ regularization ($\lambda=0.01$). 
The flattened output of $C6$ is passed as input to the session-level classification.

\emph{\textbf{RNN based Session-Level Classification: }}
The segment embeddings are extracted from the segment-level classifiers as a sequence and are passed through a Long Short-Term Memory (LSTM) based RNN model to perform the session-level classification. The input is  passed through two LSTM layers 
followed by a Dense layer with ReLU activation. Finally, the output layer with Softmax activation performs the session-level Classification. Recurrent dropout probabilities are applied to the two LSTM layers.

\vspace*{-3mm}
\subsection{Text Model}
\label{sec:text-model}
\vspace*{-2mm}
We trained a Bidirectional LSTM based HAN model shown in Fig. \ref{fig:dnn_model} to obtain a session-level classification for the text model. HAN applies the attention mechanism in two levels: word-level and sentence-level taking the hierarchical structure of a document into consideration \cite{yang-etal-2016-hierarchical}. This allows the model to learn the important words and sentences taking the context into consideration. In this work, a document corresponds to the transcribed text of a session. The embedding layer was fine-tuned for the task by allowing it to back-propagate the error from the output layer.

\vspace*{-4mm}
\subsection{Multimodal Depression Classifier}
\vspace*{-2mm}
The multimodal system (Fig. \ref{fig:dnn_model}b) is constructed with embeddings from the session-level audio classifier ($\mathbf{M_a}$) and HAN-based text classifier ($\mathbf{M_t}$). The context vector from the second LSTM layer of $\mathbf{M_a}$ and the attention-weighted sentence level context vector of $\mathbf{M_t}$ were concatenated and passed through a Dense layer with ReLu activation to perform final binary classification at the output layer. This late fusion structure helps to avoid overfitting issues that can occur as a result of the high dimensionality of input features when using early fusion. It also helps to overcome the requirement to have one-to-one correspondence between the audio segments and text sentences and allows us to create segments of different modalities independently (overlapping segments for audio and sentences for text).

\vspace*{-2mm}
\section{Experiments and Results}
\vspace*{-2mm}

\subsection{Dataset Preparation}
\vspace*{-2mm}
Similar to our previous work \cite{seneviratne21b_interspeech}, we used FS data from two databases: MD-1 \cite{MUNDT2007} and MD-2 \cite{MUNDT2012}. Both databases were collected in a longitudinal study where subjects diagnosed with MDD participated over a period of 6 and 4 weeks, respectively. For the binary classification problem, ground truth labels were determined by the bi-weekly scores provided for the clinician-rated 17-item Hamilton Depression Rating Scale (HAMD). Sessions with HAMD $>$ 7 were considered as `depressed' and sessions with HAMD $\leq$ 7 were considered as `not-depressed'. Due to the availability of 2 clinician-rated depression scores in MD-2, the agreement between the two scores in terms of the severity level was considered (see Table 1 in \cite{seneviratne21b_interspeech}). 
Originally there were 472 (35 speakers) and 753 (105 speakers) FS recordings from MD-1 and MD-2 respectively. The 140 speakers were divided into train / validation / test splits ($60:20:20$) preserving a similar class distribution in each split and ensuring that there are no speaker overlaps. For the segment-level models trained on ACFs, we segmented the audio recordings that are longer than 20s into segments of 20s with a shift of 5s. Recordings with duration less than 10s were discarded and other shorter recordings (between 10s-20s) were used as they were. Table \ref{tab:durations} summarizes the amount of speech data available after the segmentation. For the baseline model trained on openSMILE features, all audio segments were truncated at 10s (minimum length of the available audio segments) to have fixed sized inputs to the CNN. Before extracting the low-level features, segments were normalized to have a maximum absolute value of 1. 

\vspace*{-5mm}
\begin{table}[h]
    \centering
    \scriptsize
    \caption{Available Data in hours/ \# segments/ \# sessions}
    \vspace{1mm}
    \label{tab:durations}
    \begin{adjustbox}{max width = \columnwidth}
        \begin{tabular}{|c|c|c|}
        \hline
        \textbf{Database} & \textbf{Depressed} & \textbf{Not-depressed} \\ \hline
        MD-1              &  11.8 / 2131 / 111  &  2.5 / 444 /  22       \\ \hline
        MD-2              &  16.8 / 3056 / 232  & 1 / 183 / 17       \\
        \hline
        \end{tabular}
    \end{adjustbox}
    \vspace*{-2mm}
\end{table}

Before extracting GloVe embeddings for the text data, the transcribed text was preprocessed by removing punctuation,  expanding contractions, lemmatizing and removing stop words (except negation words to preserve the contextual meaning).

\vspace*{-2mm}
\subsection{Model Training}
\vspace*{-2mm}
Hyper-parameters of the models were tuned using a grid search. The ranges used were as follows: the kernel size \{(3,1), (4,1)\} and the number of output filters \{128, 64, 32, 16, 8\} of the convolutional layers, the number of hidden units of dense layers \{16, 8\}, the number of hidden units of LSTM layers \{128, 100, 64, 32\}, the dimension of the attention vectors \{128, 100, 64\} and dropout probabilities \{0.4, 0.5, 0.6, 0.7\}. 
The models were optimized using an Adam Optimizer for the Binary Cross Entropy loss. The models were trained with an early stopping criteria based on validation loss (patience 20 epochs) for a maximum of 300 epochs. A batch size was 128. All seed values were set to 1729 for training. A learning rate of $2e-5$ was used for the segment-level classifier. The session-level unimodal and multimodal classifiers were trained using an adaptive learning rate starting from $2e-4$ and it was decayed by 50\% every 10 epochs until it reached $2e-5$. To address the class imbalance issue, class weights were assigned to both training and validation splits during the training process to both the models. To evaluate the performance of the model, the Area Under the Receiver Operating Characteristics Curve (AUC-ROC), Unweighted Average Recall (UAR) and F1 scores were used.

\vspace*{-5mm}
\begin{table}[h]
    \centering
    \scriptsize
    \caption{Classification Results - Audio Model}
    \vspace{2mm}
    \label{tab:audio-results}
    \begin{adjustbox}{max width = \columnwidth}
        \begin{tabular}{|c|c|c|c|c|}
        \hline
        \textbf{Model}                                                                                         & \textbf{Features} & \textbf{AUC-ROC} & \textbf{UAR}    & \textbf{F1 (D/ND)} \\ \hline
        \textbf{Segment-Baseline}                                                                              & openSMILE         & 0.6300           & 0.5602          & 0.86/0.22                \\ \hline
        \multirow{5}{*}{\textbf{\begin{tabular}[c]{@{}c@{}}Segment-Level\\ Classifier\\ (ACFs)\end{tabular}}}  & \textbf{TV}       & \textbf{0.7408}  & \textbf{0.6961} & \textbf{0.87/0.37}       \\ \cline{2-5} 
                                                                                                               & MFCC              & 0.6031           & 0.5412          & 0.88/0.19                \\ \cline{2-5} 
                                                                                                               & FMT               & 0.5714           & 0.5688          & 0.77/0.22                \\ \cline{2-5} 
                                                                                                               & MFCC+GL           & 0.6042           & 0.5474          & 0.87/0.20                \\ \cline{2-5} 
                                                                                                               & FMT+GL            & 0.4806           & 0.5045          & 0.79/0.16                \\ \hline
        \textbf{Session-Baseline}                                                                              & openSMILE         & 0.6673           & 0.6613          & 0.87/0.35                \\ \hline
        \multirow{5}{*}{\textbf{\begin{tabular}[c]{@{}c@{}}Session-Level\\ Classifier \\ (ACFs)\end{tabular}}} & \textbf{TV}       & \textbf{0.8246}  & \textbf{0.8024} & \textbf{0.91/0.52}       \\ \cline{2-5} 
                                                                                                               & MFCC              & 0.7016           & 0.6452          & 0.85/0.32                \\ \cline{2-5} 
                                                                                                               & FMT               & 0.75             & 0.7238          & 0.88/0.42                \\ \cline{2-5} 
                                                                                                               & MFCC+GL           & 0.6794           & 0.6452          & 0.85/0.32                \\ \cline{2-5} 
                                                                                                               & FMT+GL            & 0.6552           & 0.6734          & 0.73/0.31                \\ \hline
        \end{tabular}
    \end{adjustbox}
    \vspace*{-5mm}
\end{table}

\vspace*{-2mm}
\subsection{Segment-Level to Session-Level Classification}
\vspace*{-2mm}
We trained 6 different audio-based segment-level to session-level classifiers using the ACFs derived from various feature vectors and openSMILE features (baseline). Results can be found in Table \ref{tab:audio-results}. In general, for all feature sets, there is a performance boost in the session-level classifier compared to the segment-level classifier. In both segment-level and session-level classifications, TV-based ACFs outperform the other features. TV-based ACFs yield a relative AUC-ROC improvement of 11.3\% and a relative UAR improvement of 15.3\% in session-level classification compared to segment-level classification. The AUC-ROC and UAR of session-level TV-based ACF classifier are 9.9\% and 9.8\% higher than the second best performing session-level classifier which was trained using formant-based ACFs. The chance-level F1 scores for depressed/not-depressed classes were 0.64/0.18 (segment-level) and 0.64/0.19 (session-level).

\vspace*{-5mm}
\begin{table}[h]
    \centering
    \scriptsize
    \caption{Results of Classification Using Different Modalities}
    \vspace{2mm}
    \label{tab:mm-results}
    \begin{adjustbox}{max width = \columnwidth}
       \begin{tabular}{|c|c|c|c|c|}
        \hline
        \textbf{Model}       & \textbf{Features}        & \textbf{AUC-ROC} & \textbf{UAR}    & \textbf{F1 (D) / F1 (ND)} \\ \hline
        \textbf{Audio}       & TV\_ACF                  & 0.8246           & 0.8024          & 0.91/0.52                 \\ \hline
        \textbf{Text}        & GloVe                    & 0.7802           & 0.7540          & 0.85/0.41                 \\ \hline
        \textbf{multimodal} & \textbf{TV\_ACF + GloVe} & \textbf{0.8871}  & \textbf{0.8105} & \textbf{0.92/0.55}        \\ \hline
        \end{tabular}
    \end{adjustbox}
\vspace*{-5mm}
\end{table}

\vspace*{-2mm}
\subsection{Results of multimodal Classification}
\vspace*{-2mm}
Using the best performing audio model which was trained using TV-based ACFs and HAN based text model, we trained a multimodal system that yields synergies by combining different modalities. According to Table \ref{tab:mm-results}, the multimodal system has a relative AUC-ROC improvement of 7.58\%  and 13.7\% compared to the audio model and the text model, respectively.

\begin{figure}[htb]

\begin{minipage}[b]{\linewidth}
  \centering
  \centerline{\includegraphics[scale=0.275]{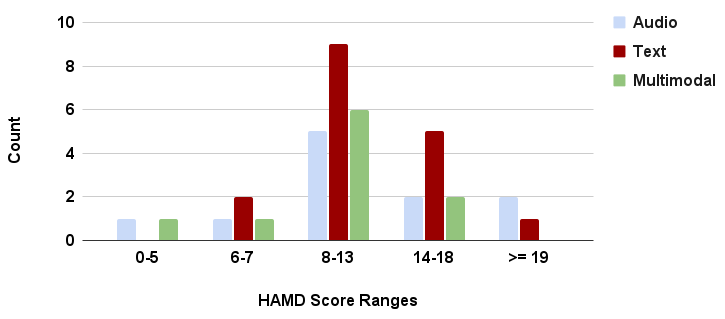}}
  \centerline{}\medskip
\end{minipage}

\vspace*{-10mm}
\caption{Distribution of HAMD scores for the mis-classified samples categorized by severity levels: not-depressed (0-5), borderline not-depressed (6-7), mild (8-13), moderate (14-18), severe ($\geq$19)}
\label{fig:hamd-misclassifications}
\end{figure}



\vspace*{-4mm}
\section{Discussion}
\vspace*{-3mm}
Results reported in Table \ref{tab:audio-results} show that TV-based ACFs are more effective compared to other feature sets in the binary depression classification task. These results support the hypothesis that TVs as a direct measure convey more distinguishing information regarding the articulatory coordination of depressed speech. Further, it is evident from the results that the performance of the session-level classifier heavily relies on the performance of the segment classifier. We observed that the segment-level TV-based classifier satisfied the constraints we derived in section \ref{sec:proof} (the recall of depressed and not-depressed classes were 58.1\% and 81.1\%, respectively). Consequently, the TV-based session-level classifier yielded better results with recall increased to 75\% and 85.5\% for the two classes respectively. 

According to Table \ref{tab:mm-results}, the audio-only model performs better than the text-only model. One possible reason for this could be errors introduced to the text transcripts by the ASR tool. We further investigated the sentence-level attention weights of the mis-classified sessions to understand the nature of the errors. Most of the misclassifications are cases where the subjects use complex sentence structures (often with mixed sentiments) in their answers. However the model seems to capture the most frequently occurring raw sentiment and not take the sentence structure into account. A common pattern is when patients use contrast in their sentences. Consider the following example. Text:  \textit{“Like I said in the beginning, I am feeling much better. I'm not feeling sad. I'm \underline{not} feeling guilty and feeling like I \underline{cannot} do anything like before”}. Ground Truth: not depressed (misclassified as depressed). Note that the first segment of the text \textit{“I am feeling much better. I'm not feeling sad. I'm not feeling guilty”} in isolation conveys a positive sentiment. The last segment of the text \textit{“feeling like I cannot do anything like before”} conveys a negative sentiment to which the model has paid higher attention. However if the underlined \textit{‘not’} is also applicable to this segment, it results in double negation and consequently the segment conveys a positive sentiment. Another common pattern is when the patients excessively use negation. Consider the following example. Text: \textit{“I \underline{do not} feel like I \underline{do not} want to do anything”}. Ground Truth: not depressed (misclassified as depressed). Note the usage of double negation which results in a complex sentence structure. In the future we plan to explore context aware embeddings which are capable of correctly processing such complex sentence structures.

For the multimodal classifier, the recall of depressed and not-depressed classes were 87.1\% and 75\%, respectively. From a clinical perspective, the model is effective in recognizing depressed subjects, reducing type-II errors (classifying a depressed subject as not-depressed). Nine out of the total of 10 mis-classified sessions by the multimodal system are a subset of the 25 mis-classified sessions by unimodal classifiers. This shows that the inter-learning among different modalities can compensate for the errors made by individual modalities. It is worth noting that when a session is incorrectly classified by both the unimodal classifiers, it is always incorrectly classified by the multimodal classifier. According to Fig. \ref{fig:hamd-misclassifications}, it can be seen that the multimodal classifier is able to correctly classify the depressed sessions in the severe category which were incorrectly classified by the unimodal classifiers.

\vspace*{-3mm}
\section{Conclusion}
\vspace*{-3mm}
We presented a multimodal system which utilizes audio data from two different depression databases and text data obtained by ASR. The proposed system performs better compared to unimodal classifiers. The robustness of the TV-based ACFs is evident from the performance of the audio-only models trained using a comprehensive set of features. We established the constraints that need to be satisfied by a segment-level classifier in order to yield a stronger session-level classifier. In the future we plan to incorporate other linguistic features to improve the text model and investigate the performance of TV-based ACFs in the depression severity score prediction task. 

\bibliographystyle{IEEEbib}
\bibliography{references}

\end{document}